\documentclass[aps,pre,twocolumn,showpacs,groupedaddress]{revtex4-1}

\usepackage[francais]{babel}
\usepackage[latin1]{inputenc}
\usepackage[T1]{fontenc}
\usepackage{graphicx}
\usepackage{amssymb}
\usepackage{amsmath}
\usepackage{wasysym}
\usepackage[bbgreekl]{mathbbol}
\usepackage{lettrine}
\usepackage{color}

\def \deg {\,^{\circ}}

\begin{document}

\title{Local rheological measurements in the granular flow around an intruder}

\author{A. Seguin$^{1,2}$, C. Coulais$^{1,2,3}$, F. Martinez$^{1,2}$, Y. Bertho$^1$ and P. Gondret$^1$}

\affiliation{$^1$ Laboratoire FAST, Univ. Paris-Sud, CNRS, Université Paris-Saclay, F-91405, Orsay, France}
\affiliation{$^2$SPEC, CEA, CNRS, Université Paris-Saclay, CEA Saclay F-91191 Gif-sur-Yvette, France}
\affiliation{$^3$Kamerlingh Onnes Lab, Universiteit Leiden, Postbus 9504, 2300 RA Leiden, The Netherlands}

\begin{abstract}
The rheological properties of granular matter within a two-dimensional flow around a moving disk is investigated experimentally. Using a combination of photoelastic and standard tessellation techniques, the strain and stress tensors are estimated at the grain scale in the time-averaged flow field around a large disk pulled at constant velocity in an assembly of smaller disks. On the one hand, one observes inhomogeneous shear rate and strongly localized shear stress and pressure fields. On the other hand, a significant dilation rate, which has the same magnitude as the shear strain rate, is reported. Significant deviations are observed with local rheology that justify the need of searching for a non-local rheology.

\end{abstract}


\maketitle

\section{Introduction}

The description of the flow properties of granular material remains a true practical and fundamental challenge. This challenge comes from the inherent non linear interactions betweens the grains and from the small separation of length scales. As a matter of fact, no (attractive) force exists between two grains when the contact is broken and the flow scale is often not much larger than the elementary grain scale. Many tools have been developed over the past decades to tackle this issue and extract the kinematic properties of grain flows, underlying the open problem of defining a single rheology for granular matter. In the past decade, a local rheology emerges based on a unique relation of the friction coefficient $\mu$ with the so-called inertial number $I$ corresponding to the shear rate made dimensionless with confining pressure and density \cite{daCruz2005,Jop2006}. This rheology works quite well by rescaling various experimental and numerical data into a consistent picture for parallel flows -- such as Couette flows or flows down inclines -- or weakly non parallel flows in a wide range of flow rates \cite{daCruz2005,GDRMidi2004,Cortet2009}. Nevertheless, this local description falls short of describing non-parallel flows where the streamlines are far from parallel and also quasi-static regimes close to the ``liquid-solid'' transition \cite{Koval2009}. Non-local effects, where the stress not only depends on the strain rate but also on its spatial variations, have been recently reported both experimentally and numerically \cite{Reddy2011,VanHecke2010b,Bouzid2013}. These effects may arise from plastic rearrangements inducing long range correlations and cooperative behaviors. Non-local modeling for granular flows is thus now considered to extend the $\mu(I)$ rheology by the introduction of a diffusive term~\cite{Bouzid2013,Henann2013,Henann2014,Kamrin2015}. This captures well the average flow fields reported in some numerical simulations and experiments where the flow is observed to be strongly inhomogeneous. Such developments were inspired by kinetic theory, where non-locality arises from the diffusion of a granular temperature into a granular flow \cite{Bocquet2001,Seguin2011}.

In this paper, we present quantitative measurements in the two-dimensional flow of small photoelastic disks around a larger intruder disk, where the local strain rate and the local stresses are recorded simultaneously. This allows one to test the local rheology framework in the case of a strongly non-parallel flow. The four invariants of the local strain rate tensor and local stress tensor reveal that the flow is strongly localized around the intruder. A tentative rheological analysis sheds light on a significant amount of dilation and some spatial variations of the friction coefficient that bring out new challenges for modeling.

\section{Experimental setup}

\begin{figure}[b]
\centering
\includegraphics[width=\linewidth]{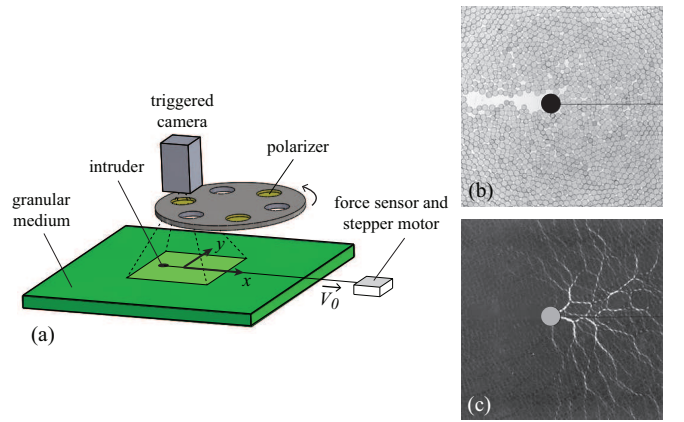}
\caption{(Color online)~(a)~Sketch of the experimental setup. (b)~Snapshot of the intruder moving from left to right through the granular packing. (c)~Same image obtained between crossed polarizers and displaying transient chains of contact forces.}
\label{Fig01}
\end{figure}

The experiments consist in pulling an intruder disk at constant velocity $V_0$ into a two-dimensional assembly of about $8\times 10^3$ smaller photoelastic disks in a set-up sketched in Fig.~\ref{Fig01}(a) which is adapted from \cite{Coulais2012}. This granular assembly is a bidisperse mixture of disks of diameters 4 and 5\,mm to avoid any possible crystallization, and made of polyurethane (density $\rho_s=1280$\,kg/m$^{3}$) which is photoelastic. These disks are placed in between two horizontal glass plates with a gap thickness of 4\,mm, slightly larger than the disk height $h=3.2$\,mm, and closed by lateral walls that form a square area of $L_x\times L_y = 400\times 400$\,mm$^2\simeq 90\times 90\,d_g^2$, where $d_g=4.5$\,mm is the mean diameter of the granular assembly. By varying the number of disks, the global packing fraction $\phi_0$ of the model granular medium has been varied in the range $0.76\leq\phi_0\leq0.80$ that corresponds to a dense packing but remains always significantly smaller than the jamming point $\phi_J\simeq 0.83$ \cite{Coulais2012}. The larger intruder disk of diameter $d=16$\,mm corresponding to about $4d_g$ is pulled in the $x$-direction with a steel wire attached to a linear stepper motor at a constant velocity $V_0$ in the range $10^{-4}\leq V_0\leq 3$\,mm~s$^{-1}$. In order to prevent the intruder from tilting and to ensure minimal interaction between the wire and the grains, the steel wire is welded on the top face of intruder and we designed the intruder slightly taller than the disks. As a result, the overhead wire hardly sags and has minimal contact with the grain assembly. Since the wire is very stiff, the intruder does not tilt more than a few degrees. The steel wire is essentially inextensible and ensures that the pulling device does not deform when the intruder travels across the medium. A piezoelectric sensor records the time evolution of the drag force $F(t)$ exerted by the granular medium onto the intruder disk during its displacement. Once the intruder placed in the granular medium at one side of the cell at its initial position $(X_i, Y_i)\simeq (-0.4L_x, 0)$ and before its pulling motion to the other side of the cell towards its final position $(X_f, Y_f)\simeq (0.4L_x, 0)$ close to the front wall, the whole packing is first homogenized by horizontal vibrations of the bottom plate in the transverse $y$-direction. All experiments presented here were performed at constant volume with a fixed cell size (constant $\phi_0$). During the intruder displacement, snapshots are recorded using a high resolution camera ($2048\times 2048$ pixels) at a frame rate $f$ such as $V_0/f < d_g/10$. To avoid any possible boundary effect in the results, images and subsequent data analysis are restricted to a central square region of interest of size $160\times 160$\,mm$^2 \simeq L_x/2.5 \times L_y/2.5$, leading thus to a spatial resolution of about $0.1$\,mm $\simeq d_g/45$. Each grain is thus composed by about $1.6 \times 10^3$ pixels in the images. Since the forces at stake are of the order of $0.1$~N, the in-plane compression is smaller than $0.01\%$, so that we consider a constant diameter in the detection algorithm. The granular packing is backlit with a large uniform source of linear polarized light and a switchable circular analyzer is mounted on a stepper motor synchronized with the picture acquisition. Such a device gives one alternated cross-polarized and direct light pictures at the effective frame rate of $f/2$. Assuming little change in the packing between two successive pictures, one combines structural information extracted from the direct light pictures [Fig.~\ref{Fig01}(b)] and stresses measurements extracted from the cross-polarized pictures [Fig.~\ref{Fig01}(c)] to quantify stress and strain tensors at the grain scale. Considering the flow geometry, the results will be presented in polar coordinates ($r, \theta$) centered on the intruder, with $\theta=0$ in the $x$-direction of the intruder motion.

\section{Image analysis}

For each direct image, we extract the position of each grain including the intruder. By tracking each grain between two successive images, we compute its velocity from its displacement. To limit noise, the velocity gradient tensor $\boldsymbol{\nabla}\mathbf{v}$ at the grain location is not calculated by differentiation but by integration along a closed contour around the grain \cite{Drescher1972,Bonelli2012}:
\begin{equation}
\boldsymbol{\nabla}\textbf{v}=\frac{1}{S}\oint \mathbf{v} \otimes \mathbf{n}~dl,
\label{Eq:1}
\end{equation}
where $S$ is the surface delimited by the corresponding contour, $dl$ is the elementary displacement on this contour for which the outward normal unit vector is $\mathbf{n}$, and $\otimes$ denotes the tensor product. Here, as each grain is surrounded by a finite and discrete number of neighbors, we compute the integral of Eq.~(\ref{Eq:1}) along the edges of its Vorono\"i cell. The strain rate tensor $\mathbf{G}$ is then deduced for each grain by
\begin{equation}
\mathbf{G} = \frac{1}{2}\left(\boldsymbol{\nabla}\mathbf{v} + \boldsymbol{\nabla}\mathbf{v}^{T}\right),
\end{equation}
where the exponent $^T$ refers to the transpose matrix. The strain rate tensor is finally interpolated along a cartesian grid of step $dx = dy=0.4$\,mm $\simeq d_g/10$.

Another important part of the image processing is the computation of the stress tensor $\boldsymbol{\sigma}$. From each direct image, the Delaunay triangulation and the Vorono\"i tessellation are constructed to extract the neighbor network and contact network of each grain at each time $t$. From the cross polarized image at the same given time $t$ (more exactly that only differs by $1/f$), we determine then the contact forces between grains by calculating the spatial gradient of the intensity inside triangular sectors defined by one particle center and the corresponding side of the Vorono\"i cell \cite{Coulais2014}. The stress tensor for each grain is then computed as \cite{Drescher1972,Bi2011}
\begin{equation}
\boldsymbol{\sigma}=\frac{4}{\pi d_i^2}\sum_{i\neq j} \mathbf{f}_{ij} \otimes\mathbf{r}_{ij},
\end{equation}
where $\mathbf{f}_{ij}$ is the contact force between the two grains $i$ and $j$ in contact and $\mathbf{r}_{ij}$ is the vector of modulus $d_i/2$ pointing from the center of grain $i$ of diameter $d_i$ towards the center of grain $j$. The stress tensor $\boldsymbol{\sigma}$ is then interpolated on the same Cartesian grid as for the strain rate tensor $\mathbf{G}$. Note that the present 2D stress is a force per unit length and not a force per unit area as the real stress $\boldsymbol{\sigma}/h$ would be.

Finally both strain rate and stress tensors $\mathbf{G}$ and $\boldsymbol{\sigma}$ are decomposed into an isotropic part and a deviatoric part \cite{Landau1986}:
\begin{equation}
\mathbf{G}= \dot{\varepsilon} \mathbf{I} + \mathbf{G_d},
\label{Eq:strain}
\end{equation}
\begin{equation}
\boldsymbol{\sigma}= -p \mathbf{I} + \boldsymbol{\sigma}_\mathbf{d},
\label{Eq:stress}
\end{equation}
where $\mathbf{I}$ is the unit tensor, $\dot{\varepsilon}=\frac{1}{2}\mathrm{tr}(\mathbf{G})=\frac{1}{2}\textrm{div}\,\mathbf{v}$ is the dilation rate of the 2D grain flow, $\mathbf{G_d}$ is the shear rate tensor, $p$ is the pressure and $\boldsymbol{\sigma}_\mathbf{d}$ is the shear stress tensor. In the following, we shall present the results using only scalar quantities corresponding to the invariants of the strain rate tensor $\mathbf{G}$ and the stress tensor $\boldsymbol{\sigma}$. The first invariants are the dilation rate $\dot{\varepsilon}$ and the pressure $p$ and the second invariants corresponding to the deviatoric parts of the two tensors are the shear rate $\dot{\gamma}$ and the shear stress $\tau$ as named usually in rheology \cite{Andreotti2013} and defined as:
\begin{equation}
\dot{\gamma}=\left(\frac{1}{2}\mathrm{tr}(\mathbf{G}_\mathbf{d}^2)\right)^{1/2},
\end{equation}
\begin{equation}
\tau=\left(\frac{1}{2}\mathrm{tr}(\boldsymbol{\sigma}_\mathbf{d}^2)\right)^{1/2}.
\end{equation}
The drag force $F$ on the intruder can be inferred from the image analysis described above by the integration of the local stress tensor around the intruder:
\begin{equation}
F=-\oint \boldsymbol{\sigma}\mathbf{n}\cdot \boldsymbol{e_x}\, dl.
\label{Eq:force}
\end{equation}
For the image analysis, we leave out the images where the intruder is too close to the edges of the image and only focus on the images where the intruder is in a central zone ($-60\leqslant X \leqslant 60$\,mm, \emph{i.e.} $-L_x/7 \lesssim X \lesssim L_x/7$). This allows one to have local information in the granular medium up to at least $22$\,mm $\simeq 5d_g$ away from the intruder center. This should be a large enough zone of analysis as the radial extension of the velocity field perturbation is expected to be here of about $2d_g+d/4 \simeq 13$\,mm $\simeq 3d_g$ \cite{Seguin2013}.

\begin{figure}[t]
\centering
\includegraphics[width=\linewidth]{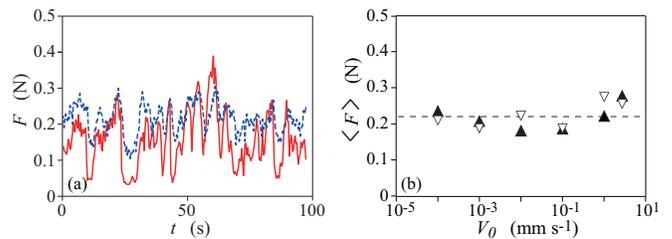}
\caption{(Color online)~(a)~Instantaneous drag force $F(t)$ on the intruder obtained either (--~--)~by the force sensor or (---)~by the image analysis from the photoelastic images for a intruder of diameter $d=16$\,mm in motion at the velocity $V_0=10^{-2}$\,mm~s$^{-1}$ in a 2D granular medium of packing fraction $\phi_0=0.76$. (b)~Time-averaged drag force $\langle F\rangle$ as a function of $V_0$. Data from ($\triangledown$)~force sensor and ($\blacktriangle$)~photoelastic images  together with  (--~--)~the mean value$\langle F\rangle=0.22$\,N averaged over the explored velocity range $10^{-5} < V_0 < 10$ mm/s.}
\label{Fig02}
\end{figure}

The time evolution of the drag force $F(t)$ extracted from the image analysis [Eq.~(\ref{Eq:force})] is reported in Fig.~\ref{Fig02}(a) together with the global measurements from the piezoelectric sensor. Both show strong fluctuations that are undoubtedly associated with the creation/breaking of force chains inside the granular material but no significant slow variations. The fluctuations of these two signals are rather well time-correlated even if their amplitudes are different which may be related to the very different measurement technics. As a matter of fact, the force sensor measures not only the drag on the intruder but also the friction of the intruder with the bottom and the possible friction of the wire attached to the intruder. By contrast, the photoelastic technique only measures intergrain contact forces. Anyway, the absence of slow time variations suggests that a stationary regime is reached where the force is constant on average, in agreement with previous observations for the motion of objects in a granular material \cite{Albert1999,Albert2001,Stone2004,Seguin2011,Seguin2013}. In the following, we only focus on time-averaged quantities along the intruder motion in the central part of the cell. Figure~\ref{Fig02}(b) shows that the mean drag forces $\langle F\rangle$ obtained from the force sensor and from the photoelastic signal are very close to each other whatever the pulling velocity $V_0$, which gives confidence in our local image analysis. For $V_0$ ranging within about five decades, $\langle F(t)\rangle$ is roughly constant, in agreement with the experiments mentioned above.

In the next section, we will focus on the time-averaged local quantities that are extracted from image analysis of both direct and polarized images. These different quantities characterizing the granular flow will be normalized by the natural length scale $d_g$ and time scale $d_g/V_0$. As there is no external stress applied on the lateral boundaries and no ``hydrostatic pressure'' from gravity in our configuration, the typical stiffness $k=(\pi/4)Eh$ of the contact between two cylinders~\cite{Popov2010} will be taken as the stress scale, where $E$ is the effective Young modulus of the disks of height $h$ with here the value $k \simeq 1$\,N mm$^{-1}$.

\section{Local measurements}

\begin{figure}[t]
\centering
\includegraphics[width=\linewidth]{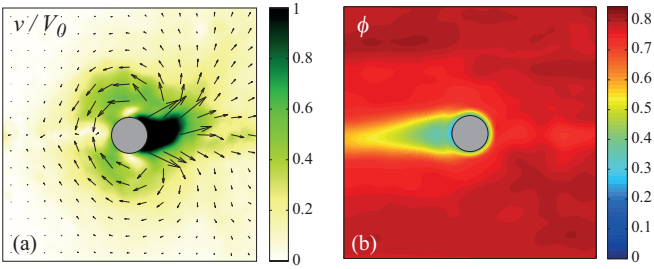}
\caption{(Color online)~(a)~Velocity field around an intruder disk of diameter $d =16$\,mm moving from left to right at the velocity $V_0=2.7$\,mm~s$^{-1}$ into a granular medium of global packing fraction $\phi_0 = 0.76$. Vectors indicate the grain velocity direction and the colormap encodes the modulus of the dimensionless velocity $v/V_0$. (b)~Corresponding local packing fraction $\phi$.}
\label{Fig03}
\end{figure}

A typical velocity field of the granular medium around the moving intruder is displayed in Fig.~\ref{Fig03}(a). As expected from previous kinematic description \cite{Seguin2011,Seguin2013}, the velocity perturbation induced by the intruder motion is found to be strongly localized around it. The corresponding local packing fraction $\phi$ is displayed in Fig.~\ref{Fig03}(b): $\phi$ is observed to be quite homogeneous in space and close to the global value $\phi_0$ except in a very narrow crown upstream of the intruder and in an elongated triangular wake downstream where $\phi$ is significantly smaller. This downstream wake has been reported in detail by \cite{Kolb2013} in similar 2D experiments and follows directly from the absence of confining pressure on the cell boundaries or gravity, as the present setup is horizontal.
\begin{figure}[t]
\centering
\includegraphics[width=\linewidth]{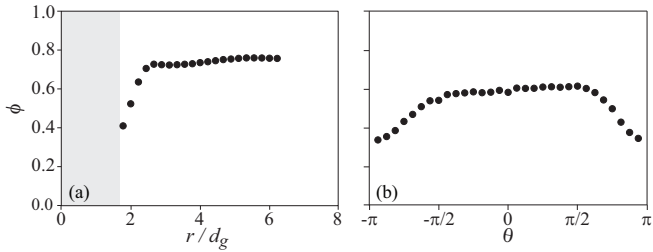}
\caption{(a)~Local packing fraction $\phi$ as a function of the radial position $r/d_g$ from the intruder along the azimuthal direction $\theta=(10\pm 3)\deg$. (b)~Local packing fraction $\phi$ as a function of the azimuthal position $\theta$ for the radial position $r=(11\pm 1)$\,mm (\emph{i.e.} $r/d_g \simeq 2.4\pm 0.2$ from the intruder). Data are obtained for $d=16$\,mm, $V_0=10^{-2}$\,mm~s$^{-1}$, and $\phi_0=0.76$.}
\label{Fig04}
\end{figure}

The radial and azimuthal profiles of $\phi$ are displayed in Figs.~\ref{Fig04}(a) and (b) along one azimuthal or radial position respectively. The data points have been obtained by digitizing the 2D interpolated grids onto annular sectors $(r-dr, r+dr),(\theta-d\theta, \theta+d\theta)$, with $dr=d_g/8$ and $d\theta=\pi/32$ Figure~\ref{Fig04}(a) shows that $\phi$ is quasi-constant and very close to $\phi_0=0.76$ upstream of the intruder except in a very narrow zone of size one grain diameter close to the intruder where $\phi$ decreases significantly and abruptly with $r$. The azimuthal profile close to the intruder [Fig.~\ref{Fig04}(b)] shows that there is no significant variation of $\phi$ around the intruder in a large azimuthal range ($-\pi/2 \lesssim\theta \lesssim \pi/2$) except in the downstream wake observed in Fig.~\ref{Fig03}(b).

\begin{figure}[t]
\centering
\includegraphics[width=\linewidth]{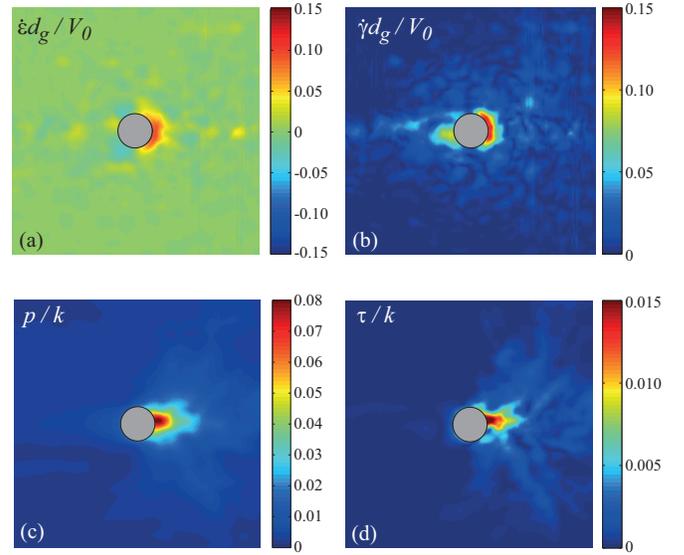}
\caption{(Color online)~Maps of the four dimensionless scalar invariants of the (a,b) strain rate tensor and (c,d) stress tensor around an intruder of diameter $d=16$\,mm moving from left to right at the velocity $V_0=2.7$\,mm~s$^{-1}$ through a granular packing of global packing fraction $\phi_0=0.76$: (a)~dilation rate $\dot{\varepsilon}d_g/V_0$, (b)~shear rate $\dot{\gamma}d_g/V_0$, (c)~pressure $p/k$, and (d)~shear stress $\tau/k$.}
\label{Fig05}
\end{figure}

The maps of the four scalar invariants $\dot{\varepsilon}$, $\dot{\gamma}$, $p$ and $\tau$ extracted from the strain rate tensor and stress tensor respectively are displayed in Fig.~\ref{Fig05}. All these quantities vary significantly upstream of the intruder with some differences. Concerning the strain rate field [Fig.~\ref{Fig05}(a)], one observes a zone of high positive $\dot{\varepsilon}$ extending about one intruder diameter upstream and in a quite large azimuthal extension ($-\pi/2 \lesssim\theta \lesssim \pi/2$), with thus a ``banana shape''. The shear rate $\dot{\gamma}$ in Fig.~\ref{Fig05}(b) displays a similar banana shape with a radial upstream extension of about half the intruder diameter and a similar azimuthal extension. This correlation between $\dot{\varepsilon}$ and $\dot{\gamma}$ is related to the fact that dense granular media are known to dilate upon transient shear \cite{Tighe2014,Coulais2014b}. Finally the stress fields [Figs.~\ref{Fig05}(c-d)] differ from the strain fields. A zone of high pressure $p$ elongated in the upstream direction appears in front of the intruder and a zone of high shear stress $\tau$ of similar shape is also observed. There is thus a clear link between $p$ and $\tau$. The relation between stress and strain appears however less clear from these four maps.

\begin{figure}[t]
\centering
\includegraphics[width=\linewidth]{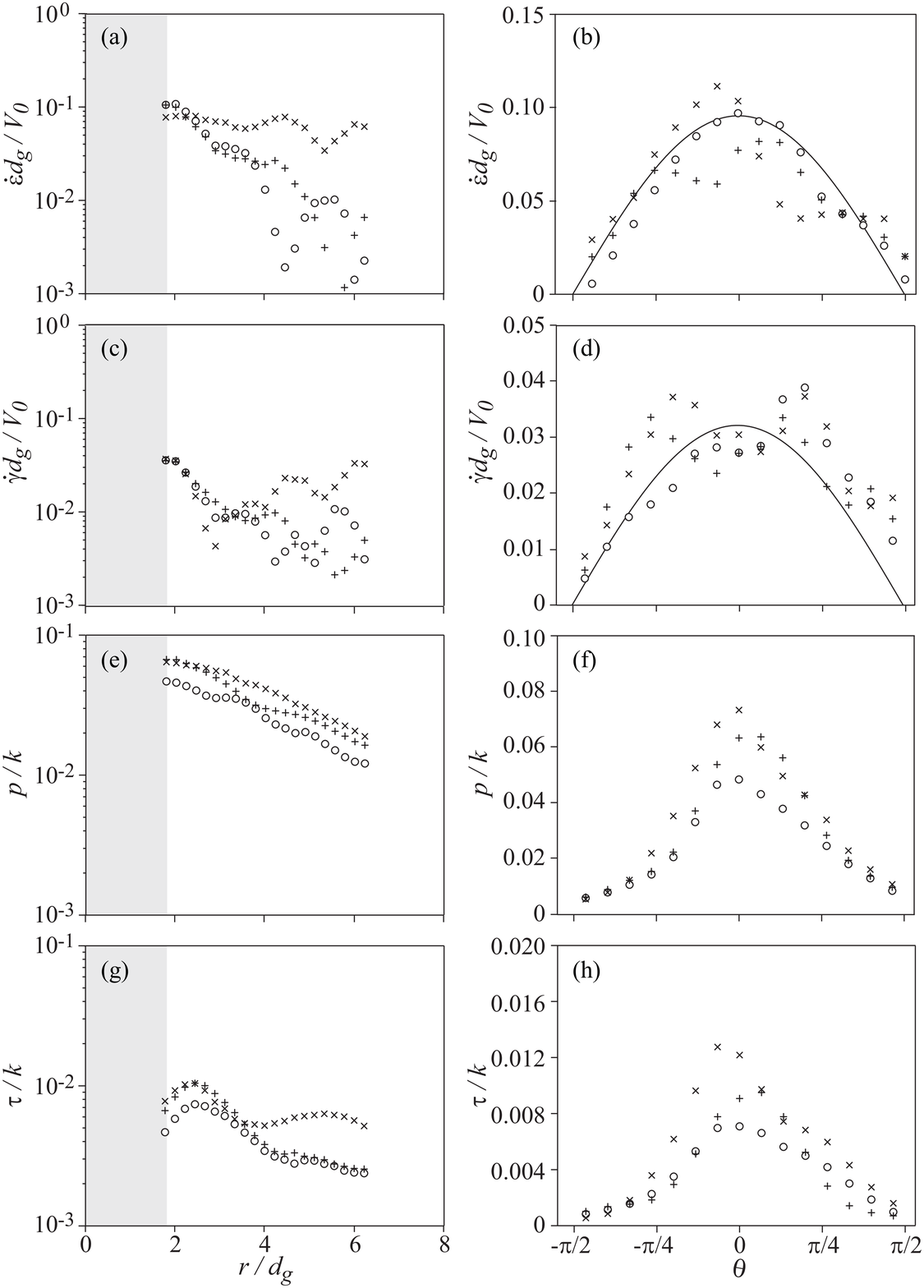}
\caption{Variations with the normalized radial distance $r/d_g$ along the azimuthal direction $\theta=(10\pm 3)\deg$, and with $\theta$ for $r =(11\pm 1)$\,mm of the four dimensionless scalar invariants: (a,b)~dilation rate $\dot{\varepsilon}d_g/V_0$, (c,d)~shear rate $\dot{\gamma}d_g/V_0$, (e,f)~pressure $p/k$, and (g,h)~shear stress $\tau/k$, for $\phi_0=0.76$ and $V_0=10^{-4}$ ($\times$), $10^{-2}$ ($\circ$), and 1 mm~s$^{-1}$ ($+$). (---)~Cosine function as a guideline for the eye. The grey zone corresponds to the intruder extension up to $r=d/2=8$\,mm.}
\label{Fig06}
\end{figure}

To quantify more precisely the spatial variations of the strain and stress tensors, the radial and azimuthal profiles of the normalized four scalar invariants are displayed in Fig.~\ref{Fig06} for the same global packing fraction ($\phi_0=0.76$) but three intruder velocities $V_0$ ranging over 4 decades. As previously, the data points have been obtained by digitizing the 2D interpolated grids onto annular sectors $(r-dr, r+dr),(\theta-d\theta, \theta+d\theta)$, with $dr=d_g/8$ and $d\theta=\pi/32$. The rather good collapse observed in Figs.~\ref{Fig06}(a-d) of the three curves for the three velocities shows that $\dot{\varepsilon}$ and $\dot{\gamma}$ scales with $V_0$ as expected. By contrast, Figs.~\ref{Fig06}(e-g) confirm that $p$ and $\tau$ do not depend significantly on $V_0$ which is consistent with the fact that the measured drag force $\langle F\rangle$ does not depend significantly on $V_0$ in Fig.~\ref{Fig02}(b). In the lin-log plots of Figs.~\ref{Fig06}(a,c,e,g), the strong monotonic decrease of $\dot{\varepsilon}$, $\dot{\gamma}$ and $p$ with increasing radial distance from the intruder is in agreement with the exponential radial decrease already reported for $\dot{\gamma}$ \cite{Seguin2011}. By contrast, $\tau$ displays a non-monotonic evolution with a first increasing and then decreasing evolution with radial increasing distance from the intruder surface. Figures~\ref{Fig06}(b,d,f,h) show that the four invariants have a maximal value in the direction of motion ($\theta=0$) but quite different variation with $\theta$. The variations of $\dot{\varepsilon}$ and $\dot{\gamma}$ are close to a cosine shape. Such a cosine shape has been already observed for $\dot{\gamma}$ in the granular flow around a cylinder in vertical penetration motions \cite{Seguin2011,Seguin2013} which is thus a consistent result. The $\theta$ variations of $p$ and $\tau$ are in contrast far from a cosine. The present observed strong localization of $p$ and $\tau$ with both $r$ and $\theta$ may explain the very small length scale reported previously for the force interaction of a moving object with both bottom \cite{Stone2004} or lateral \cite{Seguin2008} walls. Note that for the lower speed experiment ($V_0=10^{-4}$\,mm/s) the radial variations of $\dot{\epsilon}$ and $\tau$ in Figs.~\ref{Fig06}(a) and (g) are quite different from the others, which will be discussed in the next Section.

\section{Rheological curves}

\begin{figure}[t]
\centering
\includegraphics[width=\linewidth]{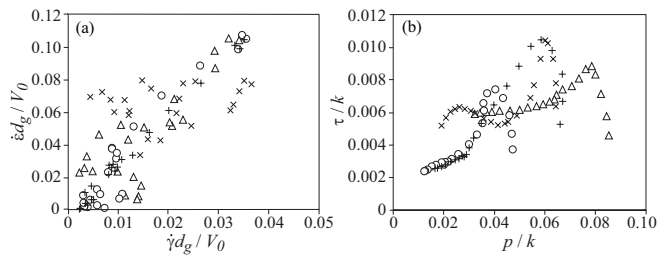}
\caption{(a)~Dimensionless dilation rate $\dot{\varepsilon}d_g/V_0$ as a function of shear rate $\dot{\gamma}d_g/V_0$ and (b)~dimensionless shear stress $\tau/k$ as a function of pressure $p/k$ for an intruder of diameter $d=16$\,mm. Same data points as in Figs.~\ref{Fig06}(a, c, e, g) together with ($\triangle$) for $\phi_0=0.80$ and $V_0=10^{-2}$\,mm~s$^{-1}$.}
\label{Fig07}
\end{figure}

Let us now look at the relations between the four scalar invariants $\dot{\varepsilon}$, $\dot{\gamma}$, $p$, and $\tau$ to search for the existence of a possible local constitutive law for the granular medium in the present non-parallel flow around a moving intruder. Despite some dispersion, Fig.~\ref{Fig07}(a) shows that the dilation rate $\dot{\varepsilon}$ increases with the shear rate $\dot{\gamma}$. This correlation between $\dot{\varepsilon}$ and $\dot{\gamma}$ is reminiscent of the so-called Reynolds dilatancy \cite{Reynolds1885}, where dilation usually increases quadratically with shear strain $\varepsilon\sim\gamma^2$ \cite{Tighe2014}. With also some dispersion, Fig.~\ref{Fig07}(b) shows that the shear stress $\tau$ increases with the pressure $p$ which justifies the introduction and use of a friction coefficient. Note that we ensure that the principal directions of the shear rate and shear stress tensors are nearly parallel in the upstream region where a local dilatancy coefficient $D$ and a friction coefficient $\mu$ will be computed, in agreement with the requirements pointed out by \cite{Cortet2009}.

\begin{figure}[t!]
\centering
\includegraphics[width=\linewidth]{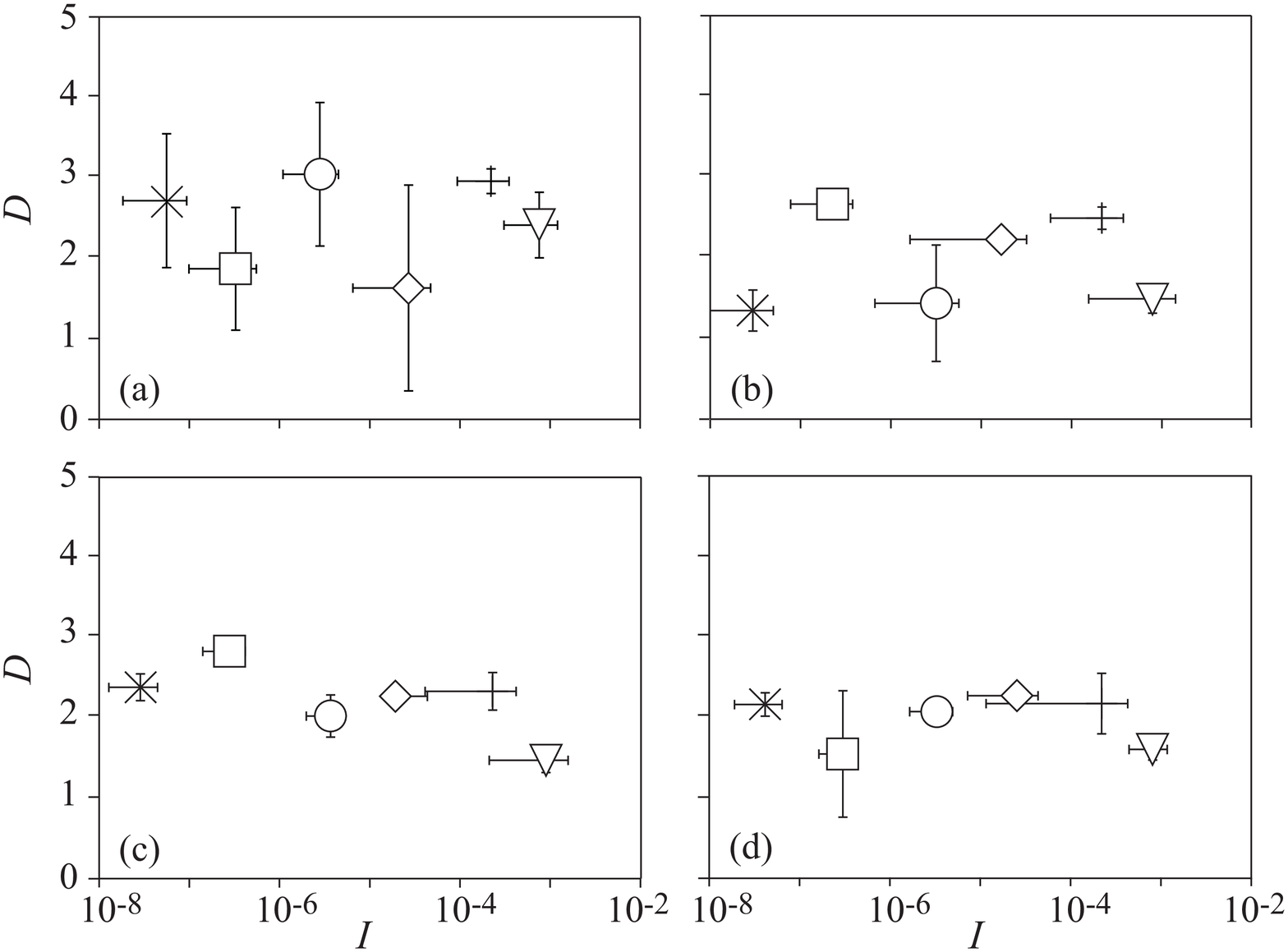}
\caption{Local dilatancy coefficient $D$ as a function of the inertial number $I$ with an average over the radial extension $r/d_g \leq 6$ along four azimuthal directions: (a)~$\theta=10\pm 3 \deg$, (b)~$\theta=45\pm 3 \deg$, (c)~$\theta=60\pm 3 \deg$, and (d)~$\theta=80\pm 3 \deg$. Same data symbols as in Fig.~\ref{Fig06} together with $V_0=10^{-3}$ ($\square$), $10^{-1}$ ($\diamond$), and $2.7$\,mm~s$^{-1}$ ($\triangledown$), for $\phi_0=0.76$.}
\label{Fig08}
\end{figure}
\begin{figure}[t!]
\centering
\includegraphics[width=\linewidth]{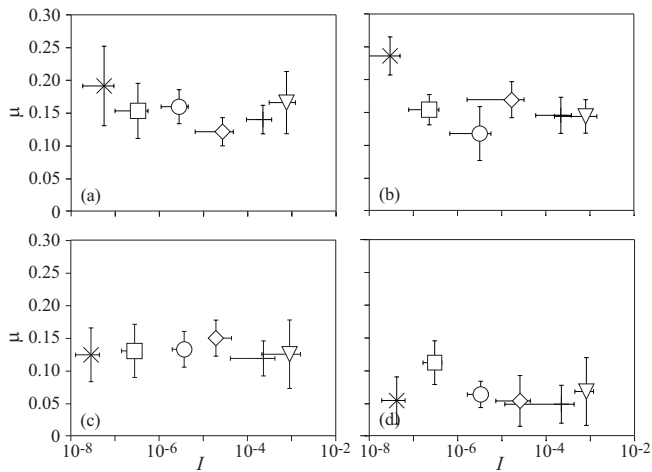}
\caption{Local friction coefficient $\mu$ as a function of the inertial number $I$ with an average over the radial extension $r/d_g \leq 6$ along four azimuthal directions: (a)~$\theta=10\pm 3 \deg$, (b)~$\theta=45\pm 3 \deg$, (c)~$\theta=60\pm 3 \deg$, and (d)~$\theta=80\pm 3 \deg$. Same data symbols as in Fig.~\ref{Fig08}.}
\label{Fig09}
\end{figure}

In order to test the possible local rheology, we now look at both relations of the dilatancy coefficient $D=\dot{\varepsilon}/\dot{\gamma}$ and the friction coefficient $\mu=\tau/p$ with the inertial number $I=\dot{\gamma}d_g\sqrt{\rho_s\phi/p}$ which is the ratio of the inertial time scale $d_g / \sqrt{p/\rho_s\phi}$ at the grain scale and the flow time scale $1/\dot{\gamma}$ , and which has been shown to be the relevant dimensionless number in simple parallel flows \cite{GDRMidi2004,daCruz2005,Jop2006}. The data shown in Figs.~\ref{Fig08} and \ref{Fig09} are the $D$, $\mu$ and $I$ values averaged over the radial range $2 \leqslant r/d_g\lesssim 6$ along different azimuthal directions, from the motion direction near $\theta=0$ [Figs.~\ref{Fig08}(a) and \ref{Fig09}(a)] towards the equatorial direction near $\theta= \pi/2$ [Figs.~\ref{Fig08}(d) and \ref{Fig09}(d)]. One observes that $D$ does not depend significantly on $I$ with about the value $D=2\pm 1$ that remains about the same whatever the azimuthal direction $\theta$. Concerning the local friction coefficient, one does not observe any significant increase of $\mu$ with $I$, which is due to the very low $I$ values explored here corresponding to a quasi-static regime ($I < 10^{-2}$). In some directions [$\theta \lesssim 45 \deg$ in Figs.~\ref{Fig09}(a) and (b)], one may even observe a significant decrease of $\mu$ with increasing $I$ at very small values ($I \lesssim 10^{-5}$). This is linked to the differences reported in Fig.~\ref{Fig06} for the lowest speed experiments (lowest $I$). This behavior is expected to arise from theoretical models \cite{Mills2008} but was not yet reported experimentally, and may explain the stick-slip like behavior observed for intruder dragged at very low velocities \cite{Albert2001}.

The key point is however the significant dependence observed for $\mu$ with the azimuthal direction $\theta$. Indeed, a significant decrease by a factor about 3 is observed from the direction of motion where $\mu \simeq 0.15$ [Fig.~\ref{Fig09}(a) for $\theta \simeq 10 \deg$] towards the equatorial direction where $\mu\simeq 0.05$ [Fig.~\ref{Fig09}(d) for $\theta \simeq 80 \deg$]. This strong $\theta$ variation of $\mu(I)$ means that such a local rheology fails here to describe the entire granular flow around an intruder. This experimental result is consistent with recent theoretical findings \cite{Henann2014} where the friction coefficient is found larger in front and behind the intruder and smaller on the sides. Our present result confirms experimentally these theoretical predictions and strongly suggests that non-local rheology is thus needed to fully describe non-parallel flows.

\section{Conclusion}

The present paper demonstrates the possibility of computation of the local stress tensor together with the local strain tensor in a two-dimensional granular packing of photoelastic disks from a detailed analysis of cross-polarized images. This possibility is here used to test a possible local rheology in a strongly non-parallel flow consisting in the grain flow around a larger intruder disk. The analysis of the spatial variations of the four invariants of the strain rate tensor and stress tensor reveals that the flow is strongly localized around the intruder and that the stress is very localized in the direction of motion. A positive dilation rate increasing roughly linearly with the shear rate is put in light and allows one for the determination of a dilatancy coefficient $D$ for the present stationary non-parallel granular flow. This dilatancy coefficient is essentially independent of the azimuthal direction $\theta$. By contrast, the friction coefficient $\mu$ that relates shear to normal stresses shows significant $\theta$ dependence. This spatial dependence demonstrates that local rheology may be not sufficient to describe strong non-parallel flows such as the present flow around a cylinder. While such spatial variations of $\mu$ have been recently predicted by non-local modeling in intruder geometry \cite{Henann2014}, the finite level of dilatancy $D$ of stationary flows has not been reported yet theoretically or numerically. The present finding may thus put further constraints on the formulation of continuum models by opening up novel challenge in modeling and understanding non-parallel compressible granular flows using compressible non-local rheology. Experimentally, checking for non-local rheology is challenging as the computation of Laplacian terms -- at the core of these non-local models \cite{Bouzid2013,Henann2014} -- is hampered by finite resolution and large fluctuations.

Similar experiments with an imposed vibration of the packing could be developed in order to uncouple the effect of the inhomogeneous excitement created by the pulling of the intruder and the effect of homogeneous excitement \cite{Candelier2009,Harich2011,Caballero2009,Djiksman2011}. These studies should help to a better understanding of the drag force with depth in the case of the vertical motion under gravity \cite{Peng2009,Guillard2013} and the direction of the motion in the penetration-extraction problem \cite{Hill2005,Schroter2007} in order to be extended for the understanding of the animal locomotion in sand \cite{Maladen2009}.

\acknowledgments
We are grateful to V. Padilla and C. Wiertel-Gasquet for the development of the experimental setup and we thank J. Crassous and A. Lemaitre for fruitful discussions. This work is supported by a public grant of the French National Research Agency (ANR) No. 2010-BLAN-0927-01 (STABINGRAM project) and supported by Triangle de la physique contracts No. 2011-075T and No. 2012-063T (COMIGS2D and REMIGS2D projects).

\bibliography{photo}

\end{document}